\documentclass[12pt]{article}
\usepackage{amssymb,amsmath,epsfig}

\begin{document}

\title{\bf Dynamics of Charged Plane Symmetric
Gravitational Collapse}
\author{M. Sharif \thanks {msharif@math.pu.edu.pk} and Aisha
Siddiqa
\thanks {asmth09@yahoo.com}\\
Department of Mathematics, University of the Punjab,\\
Quaid-e-Azam Campus, Lahore-54590, Pakistan.}

\date{}

\maketitle
\begin{abstract}
In this paper, we study dynamics of the charged plane symmetric
gravitational collapse. For this purpose, we discuss non-adiabatic
flow of a viscous fluid and deduce the results for adiabatic case.
The Einstein and Maxwell field equations are formulated for general
plane symmetric spacetime in the interior. Junction conditions
between the interior and exterior regions are derived. For the
non-adiabatic case, the exterior is taken as plane symmetric charged
Vaidya spacetime while for the adiabatic case, it is described by
plane Reissner-Nordstr$\ddot{o}$m spacetime. Using Misner and Sharp
formalism, we obtain dynamical equations to investigate the effects
of different forces over the rate of collapse. In non-adiabatic
case, a dynamical equation is joined with transport equation of heat
flux. Finally, a relation between the Weyl tensor and energy density
is found.
\end{abstract}


\section{Introduction}

Gravitational collapse is defined as the implosion of a celestial
body under the influence of its own gravity. It is one of the
basic processes driving evolution within galaxies, assembling
giant molecular clouds and producing stars. The study of
gravitational collapse is motivated by the fact that it represents
one of the few observable phenomena in the universe. The
gravitational collapse of a star proceeds to form compact objects
as white dwarf, neutron star or black hole primarily depending
upon the mass of the star. The end state of continual
gravitational collapse of a massive star is an important issue in
gravitation theory. According to Penrose, gravitational collapse
of a massive star gives rise to a spacetime singularity \cite{1}.

Oppenheimer and Snyder \cite{2} innovated the first mathematical
model for the description of gravitational collapse of stars. They
dealt with the dust case and concluded that gravitational collapse
of massive stars might end to form a black hole. This work of
Oppenheimer and Snyder attracted many people to explore it further.
Markovic and Shapiro \cite{3} extrapolated pioneer's work involving
positive cosmological constant and explained its effects on the rate
of collapse. Misner and Sharp \cite{4} did the same job by taking
ideal fluid and also discussed some other aspects like
thermodynamics and hydrodynamics of gravitational collapse.

The standard approach dealing with gravitational collapse problems
requires junction conditions to join the two spacetimes. Three types
of junction conditions are available in literature proposed by
Darmois \cite{5}, Lichnerowicz \cite{6} and O'Brien and Synge
\cite{7}. Bonnor and Vickers \cite{8} studied these junction
conditions and showed that the conditions suggested by Darmois and
Lichnerowicz are equivalent. Also, the O'Brien and Synge conditions
are unsatisfactory. It was concluded that Darmois junction
conditions are the most convenient and appropriate.

Most of the work in this scenario has been done using spherical
symmetric models \cite{9}-\cite{11}. Herrera et al. \cite{12} used
complicated fluid for this purpose and discussed the physical
meaning of expansion free fluid evolution. In another paper
\cite{13}, Herrera et al. explored the dynamics of dissipative
gravitational collapse considering spherical symmetry. They also
discussed applications of the results to astrophysical scenario.

The phenomenon of gravitational collapse has also been explored in
the context of cylindrical, planar and quasi-spherical symmetries
\cite{14}-\cite{19}. Kurita and Nakao \cite{20} studied the null
dust collapse for the cylindrically symmetric spacetimes. They
concluded that singularity must form at symmetry axis and also
discussed the behavior of geodesics arriving at the singularity.
Wang et al. \cite{21} worked on self-similar plane symmetric
solutions. It was found that during collapse trapped surfaces are
not formed and singularities are spacelike. Nath et al. \cite{22}
elaborated junction conditions for the quasi-spherical Szekeres
spacetime in the interior and the Riessner Nordstr$\ddot{o}$m Vaidya
spacetime in the exterior region. The time difference between the
formation of apparent horizon and central singularity was also
discussed.

Using the concept of gravitational lensing (GL), Virbhadra et al.
\cite{9a} introduced a new tool for examining naked singularities.
Gravitational lensing is the process of bending of light around a
massive object such as a black hole. Virbhadra and Ellis \cite{10a}
discussed GL by the Schwarzschild black hole. It was found that the
relativistic images guarantee the Schwarzschild geometry close to
event horizon. The same authors \cite{11a} also analyzed GL by a
naked singularity. Claudel et al. \cite{12a} proved that the
necessary and sufficient condition for the black hole to be
surrounded by a photon sphere is that a reasonable energy condition
holds. Virbhadra and Keeton \cite{13a} showed that weak cosmic
censorship hypothesis (CCH) can be examined observationally without
any uncertainty. Virbhadra \cite{14a} found that Seifert's
conjecture is supported by the naked singularities forming during
Vaidya null dust collapse. The same author developed an improved
form of CCH using GL phenomenon \cite{15a}

In a recent paper, Sharif and Abbas \cite{23} found the effects of
electromagnetic field on the gravitational collapse for perfect
fluid in the presence of cosmological constant. It was concluded
that charge increases the rate of collapse by decreasing the limit
of cosmological constant. The same authors extended this work for
$5$D collapse \cite{24} and found that the range of apparent horizon
is greater than $4$D case. Di Prisco et al. \cite{25} figured out
the consequences of charge and dissipation for spherical symmetric
gravitational collapse of a real fluid. It was assumed that heat
flow, free streaming radiation and shearing viscosity are the causes
of dissipation. The dynamical and transport equations are coupled to
observe the effects of dissipation over collapsing process.

This paper extends the work of Di Prisco et al. \cite{25} to plane
symmetry. The models exhibiting plane symmetry may swear out as
test-bed for numerical relativity, quantum gravity and contribute
for examining CCH and hoop conjecture among other important issues.
The paper is organized in the following pattern. In the next
section, the dynamics for non-adiabatic flow is discussed. The
Einstein-Maxwell field equations and junction conditions are found.
The dynamical and transport equations are obtained and then coupled.
A relation between energy homogeneity and the Weyl tensor is also
given. Section \textbf{3} describes dynamical equations for
adiabatic case. The last section summarizes the results.

\section{Dynamics of Viscous Non-adiabatic Case}

Here the dynamics for non-adiabatic flow is discussed. We formulate
dynamical and transport equations and then finally couple these
equations.

\subsection{Interior Spacetime and Some Physical
Quantities}

We consider a plane symmetric distribution of collapsing fluid
bounded by a hypersurface $\Sigma$. The line element for the
interior region has the following form
\begin{equation}\label{1}
ds^{2}_{-}=-A^2(t,z)dt^{2}+B^2(t,z)(dx^{2}+dy^{2})+C^2(t,z)dz^{2},
\end{equation}
where we have assumed co-moving coordinates inside $\Sigma$. The
interior coordinates are taken as
$\chi^{-0}=t,~\chi^{-1}=x,~\chi^{-2}=y,~\chi^{-3}=z$. It is
assumed that fluid is locally anisotropic and suffering
dissipation in the form of shearing viscosity, heat flow and free
streaming radiation. The energy-momentum tensor has the following
form
\begin{eqnarray}\label{2}
T_{ab}&=&(\mu+P_{\perp})V_{a}V_{b}+P_{\perp}g_{ab}+(P_{z}-P_{\perp})\chi_{a}\chi_{b}
+q_{a}V_{b}+V_{a}q_{b}+\epsilon\ell_{a}\ell_{b}\nonumber\\&-&2\eta\sigma_{ab},
\end{eqnarray}
where $\mu$ is the energy density, $P_{z}$ is the pressure in
$z$-direction, $P_{\perp}$ is the pressure perpendicular to
$z$-direction (i.e., $x$ or $y$ direction), $V^{a}$ four velocity of
fluid, $\eta$ is the coefficient of shear viscosity, $\chi^{a}$ is a
unit vector in $z$-direction, $q_{a}$ is the heat flux, $\epsilon$
is the radiation density and $\ell_{a}$ is a null four vector in
$z$-direction. Furthermore, these quantities satisfy the relations
\begin{eqnarray}\label{3}
V^{a}V_{a}&=&-1,\quad \chi^{a}\chi_{a}=1,\quad
\chi^{a}V_{a}=0,\nonumber\\
V^{a}q_{a}&=&0,\quad \ell^{a}V_{a}=-1,\quad\ell^{a}\ell_{a}=0.
\end{eqnarray}
Since the metric defined in Eq.(\ref{1}) is co-moving, we can take
\begin{eqnarray}\label{4}
V^{a}&=&A^{-1}\delta^{a}_{0},\quad \chi^{a}=C^{-1}\delta^{a}_{3},
\quad q^{a}=qC^{-1}\delta^{a}_{3},\nonumber\\
\ell^{a}&=&A^{-1}\delta^{a}_{0}+C^{-1}\delta^{a}_{3}.
\end{eqnarray}

In the standard irreversible thermodynamics by Eckart, we have the
following relation \cite{30}
\begin{equation}\label{4}
\pi_{ab}=-2\eta\sigma_{ab},\quad \Pi=-\zeta\Theta,
\end{equation}
where $\eta$ and $\zeta$ stand for coefficients of shear and bulk
viscosity, $\sigma_{ab}$ is the shear tensor, $\Theta$ is the
expansion, $\Pi$ is the bulk viscosity and $\pi_{ab}$ is the shear
viscosity tensor. The algebraic nature of Eckart constitutive
equations causes several problems but we are concerned with the
causal approach of dissipative variables. Thus we would not assume
(\ref{4}) rather we shall resort to transport equations of
M$\ddot{u}$ller-Israel-Stewart theory.

The non-zero component of acceleration and expansion scalar are
given by
\begin{equation}\label{5}
a_{3}=\frac{A'}{A},\quad
\Theta=\frac{1}{A}(\frac{2\dot{B}}{B}+\frac{\dot{C}}{C}),
\end{equation}
where dot and prime denote differentiation with respect to $t$ and
$z$ respectively. The non-vanishing components of shear tensor are
\begin{equation}\label{6}
\sigma_{11}=-\frac{1}{3}B^{2}F=\sigma_{22},\quad
\sigma_{33}=\frac{2}{3}C^{2}F;\quad
F=\frac{1}{A}(\frac{-\dot{B}}{B}+\frac{\dot{C}}{C}).
\end{equation}
The magnitude of the shear tensor, i.e., the shear scalar $\sigma$
is defined as
\begin{equation}\label{7}
\sigma^{2}=\frac{1}{2}\sigma^{ab}\sigma_{ab}=\frac{1}{9}F^{2}
\end{equation}
which implies that $F^{2}=9\sigma^{2}$.

\subsection{The Einstein and Maxwell Field Equations}

The energy-momentum tensor of an electromagnetic field is defined
as
\begin{equation}\label{8}
E_{ab}=\frac{1}{4\pi}(F^{c}_{a}F_{bc}-\frac{1}{4}F^{cd}F_{cd}g_{ab}),
\end{equation}
where $F_{ab}$ is the electromagnetic field tensor given by
\begin{equation}\label{9}
F_{ab}=\phi_{b,a}-\phi_{a,b},
\end{equation}
$\phi_{a}$ is four potential. The Maxwell field equations are
\begin{eqnarray}\label{10}
F^{ab}_{\quad ;b}=\mu_{0}J^{a},\quad F_{[ab;c]}=0,
\end{eqnarray}
where $\mu_{0}=4\pi$ is the magnetic permeability and $J_{a}$ is the
four current. In a co-moving frame $\phi_{a}$ and $J_{a}$ are given
by
\begin{equation}\label{12}
\phi_{a}=\phi\delta^{a}_{0},\quad J^{a}=\xi V^{a},
\end{equation}
where $\xi$ and $\phi$ are the charge density and electric scalar
potential respectively and both are functions of $t$ and $z$. The
charge conservation ${J^{a}}_{;a}=0$ gives the charge for interior
region as
\begin{equation}\label{13}
s(z)=\int^{z}_{0}\xi B^{2}Cdz.
\end{equation}

For $a=0,3$, the first of Eq.(\ref{10}) becomes
\begin{eqnarray}\label{14}
\phi''-[\frac{A'}{A}+\frac{C'}{C}-\frac{2B'}{B}]\phi'&=&\mu_{0}\xi
AC^{2}, \\\label{15}
\dot{\phi}'-[\frac{\dot{A}}{A}+\frac{\dot{C}}{C}-\frac{2\dot{B}}{B}]\phi'&=&0,
\end{eqnarray}
while for $a=1,2$, it is trivially satisfied. Also, the second of
Eq.(\ref{10}) is identically satisfied. Integration of Eq.(\ref{14})
with respect to $z$, assuming $\phi'(t,0)=0$, gives
\begin{equation}\label{16}
\phi'=\frac{\mu_{0}s(z)AC}{B^{2}}.
\end{equation}
The Taub's mass function \cite{26} for plane symmetric spacetime
can be generalized to include the electromagnetic contributions as
\begin{equation}\label{17}
m(t,z)=\frac{B}{2}(\frac{\dot{B}^{2}}{A^{2}}-\frac{B'^{2}}{C^{2}})+\frac{s^{2}}{2B}.
\end{equation}

For the interior spacetime, the Einstein field equations,
$G_{ab}=8\pi(T_{ab}+E_{ab})$, yield the following set of equations
\begin{eqnarray}\label{18}
&&8\pi(\mu+\epsilon)A^{2}+\frac{(\mu_{0}sA)^{2}}{B^{4}}=\frac{\dot{B}}{B}(\frac{2\dot{C}}{C}
+\frac{\dot{B}}{B})+(\frac{A}{C})^{2}[\frac{-2B''}{B}\nonumber\\
&&+(\frac{2C'}{C}-\frac{B'}{B})\frac{B'}{B}],\\\label{19} &&-8\pi
AC(q+\epsilon)
=-\frac{\dot{B}'}{B}+\frac{A'\dot{B}}{AB}+\frac{\dot{C}B'}{CB},\\\label{20}
&&8\pi(P_{\perp}+\frac{2}{3}\eta
F)B^{2}+(\frac{\mu_{0}s}{B})^{2}=-(\frac{B}{A})^{2}
[\frac{\ddot{B}}{B}+\frac{\ddot{C}}{C}-\frac{\dot{A}}{A}(\frac{\dot{B}}{B}\nonumber\\
&&+\frac{\dot{C}}{C})+\frac{\dot{B}\dot{C}}{BC}]+\frac{B^{2}}{C^{2}}[\frac{A''}{A}+\frac{B''}{B}
-\frac{A'}{A}(\frac{C'}{C}-\frac{B'}{B})-\frac{B'C'}{BC}],
\\\label{21}
&&8\pi(P_{z}+\epsilon-\frac{4}{3}\eta
F)C^{2}-\frac{(\mu_{0}sC)^{2}}{B^{4}}=-(\frac{C}{A})^{2}
[\frac{2\ddot{B}}{B}+(\frac{\dot{B}}{B})^{2}\nonumber\\
&&-\frac{2\dot{A}\dot{B}}{AB}]+(\frac{B'}{B})^{2}+\frac{2A'B'}{AB}.
\end{eqnarray}
In view of Eqs.(\ref{5}) and (\ref{6}), Eq.(\ref{19}) can be
written as
\begin{equation}\label{22}
4\pi(q+\epsilon)C=\frac{1}{3}(\Theta-F)'-F\frac{B'}{B}.
\end{equation}

\subsection{Junction Conditions}

Here we formulate junction conditions for the general plane
symmetric spacetime in the interior and charged Vaidya plane
symmetric spacetime in the exterior. The line element for the
exterior region is \cite{27}
\begin{equation}\label{23}
ds^{2}_{+}=(\frac{2M(\nu)}{Z}-\frac{e^{2}(\nu)}{Z^{2}})d\nu^{2}-2dZd\nu+Z^{2}(dX^2+dY^2),
\end{equation}
where $\chi^{+0}=\nu,~\chi^{+1}=X,~\chi^{+2}=Y,~\chi^{+3}=Z$. The
metric for hypersurface is defined as
\begin{equation}\label{24}
(ds^{2})_{\Sigma}=-(d\tau)^{2}+f(\tau)^{2}(dx^{2}+dy^{2}),
\end{equation}
where $w^{i}=(\tau,x,y) (i=0,1,2)$ are the intrinsic coordinates
of $\Sigma$. The equations of hypersurface in terms of interior
and exterior coordinates are
\begin{eqnarray}\label{25}
k^{-}(t,z)&=&z-z_{\Sigma}=0,\\\label{26}
k^{+}(\nu,Z)&=&Z-Z_{\Sigma}(\nu)=0,
\end{eqnarray}
where $z_{\Sigma}$ is constant. Using Eqs.(\ref{25}) and
(\ref{26}), we get the interior and exterior metrics over the
hypersurface as
\begin{eqnarray}\label{27}
(ds^{2}_{-})_{\Sigma}&=&-A^2(t,z_{\Sigma})dt^{2}+B^2(t,z_{\Sigma})(dx^{2}+dy^{2}),
\\\label{28}
(ds^{2}_{+})_{\Sigma}&=&-[(\frac{-2M(\nu)}{Z_{\Sigma}}+\frac{e^2(\nu)}{Z_{\Sigma}^2})
+\frac{2dZ_{\Sigma}}{d\nu}]d\nu^{2}\nonumber\\
&+&Z^{2}_{\Sigma}(dX^2+dY^2).
\end{eqnarray}

Now we use the junction conditions proposed by Darmois \cite{5}, the
first condition is
\begin{equation}\label{29}
(ds^{2})_{\Sigma}=(ds^{2}_{-})_{\Sigma}=(ds^{2}_{+})_{\Sigma}
\end{equation}
which yields the following equations
\begin{eqnarray}\label{30}
\frac{dt}{d\tau}&=&\frac{1}{A},\\\label{31}
Z_{\Sigma}&=&B,\\\label{32}
\frac{d\nu}{d\tau}&=&[(\frac{-2M(\nu)}{Z_{\Sigma}}+\frac{e^2(\nu)}{Z_{\Sigma}^2})
+\frac{2dZ_{\Sigma}}{d\nu}]^{\frac{-1}{2}}.
\end{eqnarray}
The second junction condition is the continuity of extrinsic
curvature (the second fundamental form)
\begin{equation}\label{33}
K_{ab}=K_{ab}^{-}=K_{ab}^{+}.
\end{equation}
The unit normal in terms of interior and exterior coordinates are
given respectively as
\begin{eqnarray}\label{34}
n^{-}_{a}=C(0,0,0,1),\quad
n^{+}_{a}=(-\hat{Z_{\Sigma}},0,0,\hat{\nu}),
\end{eqnarray}
here hat denotes differentiation with respect to $\tau$. The
surviving components of the extrinsic curvature for the interior
spacetime are
\begin{eqnarray}\label{36}
K_{00}^{-}=-[\frac{A'}{AC}]_{\Sigma},\quad
K_{11}^{-}=[\frac{BB'}{C}]_{\Sigma}=K_{22}^{-}.
\end{eqnarray}
The non-null components of the extrinsic curvature for the
exterior spacetime are given by
\begin{equation}\label{38}
K_{00}^{+}=[\frac{d^{2}\nu}{d\tau^{2}}(\frac{d\nu}{d\tau})^{-1}-
(\frac{M}{Z^2}-\frac{e^2}{Z^3})(\frac{d\nu}{d\tau})]_{\Sigma}.
\end{equation}
\begin{equation}\label{39}
K_{11}^{+}=[Z\frac{dZ}{d\tau}+(\frac{e^2}{Z}-2M)
\frac{d\nu}{d\tau}]_{\Sigma}=K_{22}^{+}.
\end{equation}

Thus the second junction condition yields the following equations
\begin{eqnarray}\label{40}
-[\frac{A'}{AC}]_{\Sigma}&=&[\frac{d^{2}\nu}{d\tau^{2}}(\frac{d\nu}{d\tau})^{-1}-
(\frac{M}{Z^2}-\frac{e^2}{Z^3})(\frac{d\nu}{d\tau})]_{\Sigma},
\\\label{41}
[\frac{BB'}{C}]_{\Sigma}&=&[Z\frac{dZ}{d\tau}+(\frac{e^2}{Z}-2M)
\frac{d\nu}{d\tau}]_{\Sigma}.
\end{eqnarray}
After some algebra, it follows that
\begin{equation}\label{42}
\quad M(\nu)\overset{\Sigma}{=}m(t,z)\quad \Leftrightarrow\quad
s\overset{\Sigma}{=}e,
\end{equation}
\begin{equation}\label{43}
q\overset{\Sigma}{=}P_{z}-\frac{4}{3}\eta
F-\frac{s^2}{2B^4}(\mu_{0}^2-1).
\end{equation}
These equations give necessary and sufficient conditions for the
matching of interior and exterior spacetimes. Equation (\ref{43})
describes a relationship between heat flux, effective pressure in
$z$-direction and charge over the hypersurface. It shows that if the
fluid has no charge then effective pressure and heat flux are equal
over the hypersurface.

\subsection{Dynamical Equations}

Here we develop equations that govern the dynamics of non-adiabatic
plane symmetric collapsing process by using Misner-Sharp formalism
\cite{4}. The proper time derivative and proper derivative in
$z$-direction are defined respectively \cite{16} as
\begin{eqnarray}\label{44}
D_{\tilde{T}}&=&\frac{1}{A}\frac{\partial}{\partial
t},\\\label{45}
D_{\tilde{Z}}&=&\frac{1}{\tilde{Z}'}\frac{\partial}{\partial z},
\end{eqnarray}
where $\tilde{Z}=B$. The velocity of the collapsing fluid is the
proper time derivative of $\tilde{Z}$ defined as
\begin{equation}\label{46}
U=D_{\tilde{T}}(\tilde{Z})=\frac{\dot{B}}{A}
\end{equation}
which is always negative. Using this expression, Eq.(\ref{17})
implies that
\begin{equation}\label{47}
E=\frac{B'}{C}=[U^{2}-\frac{2m}{B}+\frac{s^{2}}{B^{2}}]^{\frac{1}{2}}.
\end{equation}
When we make use of Eq.(\ref{45}) in (\ref{22}), we have
\begin{equation}\label{48}
4\pi C(q+\epsilon)
=E[\frac{1}{3}D_{\tilde{Z}}(\Theta-F)-\frac{F}{\tilde{Z}}].
\end{equation}
The rate of change of mass (given in Eq.(\ref{17})) with respect
to proper time is given by
\begin{equation}\label{49}
D_{\tilde{T}}m=-4\pi[(P_{z}+\epsilon-\frac{4}{3}\eta
F)U+E(q+\epsilon)]\tilde{Z}^2
+\frac{s^2U}{2\tilde{Z^2}}(\mu_{0}^2-1).
\end{equation}

This equation shows how mass is varying within the plane
hypersurface under the influence of matter variables. The first term
represents effective pressure in $z$-direction and radiation
density. When collapse takes place, this term is positive implying
that energy increases by this factor. The second term in square
brackets shows that energy is going out from the plane hypersurface
while the last term is the charge contribution. During collapse,
energy decreases due to these terms. Similarly, we calculate
\begin{equation}\label{50}
D_{\tilde{Z}}m=4\pi[\mu+\epsilon+(q+\epsilon)\frac{U}{E}]\tilde{Z}^2
+\frac{s}{\tilde{Z}}D_{\tilde{Z}}s
+\frac{s^2}{2\tilde{Z^2}}(\mu_{0}^2-1).
\end{equation}
This equation describes how different quantities influence the
mass between neighboring hypersurfaces in the fluid distribution.
The term $(\mu+\epsilon)$ indicates the effects of energy density
and radiation density. Similarly, the second term shows the amount
of heat and radiation which is getting out. The remaining two
terms represent contribution of electric charge. Integration of
Eq.(\ref{50}) yields
\begin{eqnarray}\label{51}
m&=&\int^{\tilde{Z}}_{0}4\pi[\mu+\epsilon+(q+\epsilon)\frac{U}{E}]\tilde{Z}^2d\tilde{Z}+
\frac{s^{2}}{2\tilde{Z}}+\frac{1}{2}\int^{\tilde{Z}}_{0}(\frac{s^{2}}
{\tilde{Z}^{2}})d\tilde{Z}\nonumber\\
&+&\frac{(\mu_{0}^{2}-1)}{2}\int^{\tilde{Z}}_{0}(\frac{s^{2}}{\tilde{Z}^{2}})d\tilde{Z}.
\end{eqnarray}

The dynamical equations can be obtained from the contracted
Bianchi identities $(T^{ab}+E^{ab})_{;b}=0$. Consider the
following two equations
\begin{eqnarray}\label{52}
(T^{ab}+E^{ab})_{;b}V_{a}=(T^{0b}_{;b}+E^{0b}_{;b})V_{0}=0,\\\label{53}
(T^{ab}+E^{ab})_{;b}\chi_{a}=(T^{3b}_{;b}+E^{3b}_{;b})\chi_{3}=0
\end{eqnarray}
which yield
\begin{eqnarray}\label{54}
&&\frac{(\mu+\epsilon)^{\cdot}}{A}+(\mu+2\epsilon+P_{z}-\frac{4}{3}\eta
F)\frac{\dot{C}}{AC}+2(\mu+\epsilon+P_{\perp}+\frac{2}{3}\eta F)\nonumber\\
&&\frac{\dot{B}}{AB}+\frac{(q+\epsilon)'}{AC}+\frac{2A'}{AC}(q+\epsilon)
+\frac{2B'}{BC}(q+\epsilon)=0,\\\label{55}
&&\frac{(q+\epsilon)^{\cdot}}{A}+\frac{1}{C}(P_{z}+\epsilon-\frac{4}{3}\eta
F)'+\frac{2(q+\epsilon)(BC)^{\cdot}}{ABC}+(\mu+P_{z}+2\epsilon\nonumber\\
&&-\frac{4}{3}\eta
F)\frac{A'}{AC}+2(P_{z}-P_{\perp}+\epsilon-2\eta
F)\frac{B'}{BC}-\frac{\mu_{0}^{2}ss'}{4\pi CB^4}=0.
\end{eqnarray}
The acceleration of the collapsing fluid is defined as
\begin{equation}\label{56}
D_{\tilde{T}}U=\frac{1}{A}\frac{\partial U}{\partial
t}=\frac{\ddot{B}}{A^{2}}-\frac{\dot{A}\dot{B}}{A^{3}}.
\end{equation}
Using Eqs.(\ref{21}), (\ref{44}) and (\ref{17}), we have
\begin{equation}\label{57}
D_{\tilde{T}}U=-4\pi(P_{z}+\epsilon-\frac{4}{3}\eta
F)\tilde{Z}-\frac{m}{\tilde{Z}^{2}}+\frac{s^{2}}{2\tilde{Z}^{3}}(\mu_{0}^{2}+1)+\frac{EA'}{AC}
\end{equation}
which gives the value of $\frac{A'}{A}$
\begin{equation}\label{58}
\frac{A'}{A}=\frac{C}{E}[D_{\tilde{T}}U+4\pi(P_{z}+\epsilon-\frac{4}{3}\eta
F)\tilde{Z}]+\frac{mC}{E\tilde{Z}^{2}}-\frac{Cs^{2}}{2E\tilde{Z}^{3}}(\mu_{0}^{2}+1).
\end{equation}
Substituting this value in Eq.(\ref{55}), it follows that
\begin{eqnarray}\label{59}
&&(\mu+P_{z}+2\epsilon-\frac{4}{3}\eta
F)D_{\tilde{T}}U=-(\mu+P_{z}+2\epsilon-\frac{4}{3}\eta
F)[\frac{m}{\tilde{Z}^{2}}+4\pi(P_{z}\nonumber\\
&&+\epsilon-\frac{4}{3}\eta F)\tilde{Z}
-\frac{s^{2}}{2\tilde{Z}^{3}}(\mu_{0}^2+1)]-
E^{2}[D_{\tilde{Z}}(P_{z}+\epsilon-\frac{4}{3}\eta
F)+\frac{2}{\tilde{Z}}(P_{z}\nonumber\\
&&-P_{\perp}+\epsilon-2\eta
F)-\frac{\mu_{0}^{2}sD_{\tilde{Z}}s}{4\pi \tilde{Z}^{4}}]
-E[D_{\tilde{T}}(q+\epsilon)+4(q+\epsilon)\frac{U}{\tilde{Z}}\nonumber\\
&&+2(q+\epsilon)F].
\end{eqnarray}

This equation yields the effect of different forces on the
collapsing process. It can be interpreted in the form of Newton's
second law of motion i.e., force = mass density $\times$
acceleration. The term within round brackets on LHS represents the
inertial or passive gravitational mass density. This term shows that
effective pressure, energy density and density of null fluid have
effects on mass density while heat flux and charge have no
contribution here. By equivalence principle, the round brackets
factor on RHS is taken as active gravitational mass density. The
quantities within square brackets in the first term show the
influence of effective pressure, radiation density and electric
charge on active gravitational mass. Using Eq.(\ref{51}) in
(\ref{59}), it follows that charge increases the active
gravitational mass if
\begin{equation}\label{60}
\frac{s^{2}}{2\tilde{Z}^{3}}+\frac{\mu_{0}^{2}}{2\tilde{Z}^{2}}\int^{\tilde{Z}}_{0}
(\frac{s^{2}}{2\tilde{Z}^{2}})d\tilde{Z}-\frac{s^{2}}{2\tilde{Z}^{3}}
(\mu_{0}^{2}+1)>0\Rightarrow \frac{s}{\tilde{Z}}>D_{\tilde{Z}}s.
\end{equation}
If this inequality holds then charge regeneration phenomenon
analogous to pressure regeneration occurs \cite{28}. The pressure
regeneration means that the pressure which is trying to keep
hydrostatic equilibrium by balancing gravitational attraction, at
the same time contributes to the active gravitational mass. This
implies that it promotes gravitational collapse. Otherwise, if the
above inequality is not satisfied, charge will decrease active
gravitational mass and consequently the Coulomb repulsion may
prevent the gravitational collapse.

The first term in the second square brackets is the gradient of
effective pressure in $z$-direction and radiation density. Since
this gradient is negative, it increases the rate of collapse. The
second term is due to local anisotropy of pressure, radiation
density and contribution of viscosity. If this term is positive
then it contributes to increase collapse and vice versa. The last
term depicts Coulomb repulsion that opposes gravitation implying
that it decelerates the collapsing process.

Finally, the last square brackets is entirely due to dissipation. To
see the role of $D_{\tilde{T}}q$, this equation is coupled with
causal transport equation. The consequences of
$D_{\tilde{T}}\epsilon$ have been discussed by Misner \cite{29}. The
outward flux of radiation accelerates collapsing process by
increasing gravitational force. The third term is positive as $U<0$,
so it slows down rate of collapse. The last term shows the combine
effect of viscosity and dissipation. From Eq.(\ref{59}), the
condition for hydrostatic equilibrium can be obtained by replacing
$U=q=\epsilon=\eta=0$ as
\begin{equation}\label{61}
D_{\tilde{Z}}P_{z}=-\frac{(P_{z}+\mu)}{E^2}[\frac{m}{\tilde{Z}^2}-\frac{s^2}{2\tilde{Z}^3}
(\mu_{0}^2+1)]+\frac{\mu_{0}^{2}sD_{\tilde{Z}}s}{4\pi
\tilde{Z}^{4}}-\frac{2}{\tilde{Z}}(P_{z}-P_{\perp}).
\end{equation}

\subsection{Transport Equation}

The transport equation for heat flux derived from the
M$\ddot{u}$ller-Israel-Stewart theory of dissipative fluids
\cite{30} is given by
\begin{equation}\label{62}
\tau_{0}h^{ab}V^{c}q_{b;c}+q^{a}=-\kappa
h^{ab}(T_{,b}+Ta_{b})-\frac{1}{2}\kappa
T^2(\frac{\tau_{0}V^{b}}{\kappa T^2})_{;b}q^{a},
\end{equation}
where $h^{ab}$ is the projection tensor, $\kappa$ denotes thermal
conductivity, $T$ is temperature and $\tau$ stands for relaxation
time. This equation has only one independent component
\begin{eqnarray}\label{63}
D_{\tilde{T}}q&=&-\frac{\kappa
T^2q}{2\tau_{0}}D_{\tilde{T}}(\frac{\tau_{0}}{\kappa
T^2})-q[\frac{3U}{2\tilde{Z}}+\frac{F}{2}+\frac{1}{\tau_{0}}]-\frac{\kappa
E}{\tau_{0}}D_{\tilde{Z}}T-\frac{\kappa
T}{\tau_{0}E}\nonumber\\
&\times&D_{\tilde{T}}U -\frac{\kappa
T}{\tau_{0}E}[m+4\pi(P_{z}+\epsilon-\frac{4}{3}\eta
F)\tilde{Z}^3-\frac{s^{2}}{\tilde{Z}}]\frac{1}{\tilde{Z}^2}.
\end{eqnarray}

We now couple this equation with dynamical Eq.(\ref{59}) to see the
effects of heat flux or dissipation on collapsing process. Replacing
Eq.(\ref{63}) in (\ref{59}), we obtain
\begin{eqnarray}\label{64}
&&(\mu+P_{z}+2\epsilon-\frac{4}{3}\eta F)(1-\alpha)D_{\tilde{T}}U=
(1-\alpha)F_{grav}+F_{hyd}+\frac{\kappa
E^2}{\tau_{0}}\nonumber\\
&&D_{\tilde{Z}}T+E[\frac{\kappa
T^2q}{2\tau_{0}}D_{\tilde{T}}(\frac{\tau_{0}}{\kappa
T^2})-D_{\tilde{T}}\epsilon]-Eq(\frac{5U}{2\tilde{Z}}
+\frac{3}{2}F-\frac{1}{\tau_{0}})-2E\epsilon\nonumber\\
&&(\frac{2U}{\tilde{Z}}+F),
\end{eqnarray}
where $F_{grav},~F_{hyd}$ and $\alpha$ are given by the following
equations
\begin{eqnarray}\label{65}
F_{grav}&=&-(\mu+P_{z}+2\epsilon-\frac{4}{3}\eta
F)[m+4\pi(P_{z}+\epsilon-\frac{4}{3}\eta
F)\tilde{Z}^3-\frac{s^2}{2\tilde{Z}}\nonumber\\
&\times&(\mu_{0}^2+1)]\frac{1}{\tilde{Z}^2},\\\label{66}
F_{hyd}&=&-E^2[D_{\tilde{Z}}(P_{z}+2\epsilon-\frac{4}{3}\eta
F)+\frac{2}{\tilde{Z}}(P_{z}-P_{\perp}+\epsilon-2\eta F)
-\frac{sD_{\tilde{Z}}s}{4\pi\tilde{Z}^4}],\nonumber\\\\\label{67}
\alpha&=&\frac{\kappa
T}{\tau_{0}}(\mu+P_{z}+2\epsilon-\frac{4}{3}\eta F)^{-1}.
\end{eqnarray}
The consequence of coupling transport and dynamical equations is
that the inertial and active gravitational mass densities are
affected by a factor $\alpha$ given by Eq.(\ref{67}). The
gravitational force term defined in Eq.(\ref{65}) is also affected
by $\alpha$ but the hydrodynamical forces Eq.(\ref{66}) are not
influenced by this term.

\subsection{Relation Between the Weyl Tensor and Matter
Variables}

Here we find some relationship between the Weyl tensor and matter
variables. The Weyl scalar $\mathcal{C}$ in terms of Kretchman
scalar $\mathcal{R}$, the Ricci tensor $R_{ab}$ and the Ricci
scalar $R$ is given by
\begin{equation}\label{68}
\mathcal{C}^{2}=\mathcal{R}-2R^{ab}R_{ab}+\frac{1}{3}R^{2}.
\end{equation}
The Kretchman scalar $\mathcal{R}^{2}=R^{abcd}R_{abcd}$ becomes
\begin{eqnarray}\label{70}
\mathcal{R}^{2}&=&4[\frac{2}{A^{4}B^{4}}(R^{0101})^{2}+\frac{1}{A^4C^4}(R^{0303})^{2}
+\frac{1}{B^{8}}(R^{1212})^{2}\nonumber\\
&+&\frac{2}{B^4C^4}(R^{2323})^{2}-\frac{4}{A^2B^4C^2}(R^{0113})^2].
\end{eqnarray}
The non-zero components of the Riemann tensor can be written in
terms of the Einstein tensor and mass function as
\begin{eqnarray*}
R_{0101}&=&(AB)^{2}[\frac{1}{2C^{2}}G_{33}+\frac{1}{B^{3}}(m-\frac{s^{2}}{2B})]
=R_{0202},\\
R_{0303}&=&(AC)^{2}[\frac{1}{2A^{2}}G_{00}-\frac{1}{2C^{2}}G_{33}+\frac{1}{B^{2}}G_{22}
-\frac{2}{B^{3}}(m-\frac{s^{2}}{2B})],\\
R_{1212}&=&2B(m-\frac{s^{2}}{2B}),\\
R_{1313}&=&(BC)^{2}[\frac{1}{2A^2}G_{00}-\frac{1}{B^3}(m-\frac{s^2}{2B})]=R_{2323},\\
R_{0113}&=&\frac{-B^{2}}{2}G_{03}.
\end{eqnarray*}
Substituting these values in Eq.(\ref{70}), after some algebra, we
obtain
\begin{eqnarray}\label{71}
\mathcal{R}^{2}&=&\frac{48}{B^{6}}(m-\frac{s^2}{2B})^{2}-\frac{16}{B^3}(m-\frac{s^2}{2B})
[\frac{G_{00}}{A^2}-\frac{G_{33}}{C^2}+\frac{G_{22}}{B^{2}}]\nonumber\\
&-&\frac{4}{A^2C^2}G_{03}^{2}+3[(\frac{G_{00}}{A^2})^{2}+(\frac{G_{33}}{C^2})^{2}]
+\frac{4}{B^4}G_{22}^{2}-2\frac{G_{00}G_{33}}{A^2C^2}\nonumber\\
&+&4(\frac{G_{00}}{A^2}-\frac{G_{33}}{C^2})\frac{G_{22}}{B^2}.
\end{eqnarray}

Now we calculate the remaining part of the Weyl scalars which need
Ricci tensor and Ricci scalar in terms of the Einstein tensor.
These are
\begin{eqnarray*}
R_{00}&=&A^2[\frac{G_{00}}{2A^2}+\frac{G_{33}}{2C^2}+\frac{G_{22}}{B^2}],\quad
R_{03}=G_{03},\\
R_{11}&=&\frac{B^2}{2}[\frac{G_{00}}{A^2}-\frac{G_{33}}{C^2}]=R_{22},\quad
R_{33}=C^2[\frac{G_{00}}{2A^2}+\frac{G_{33}}{2C^2}-\frac{G_{22}}{B^2}],\\
R&=&\frac{G_{00}}{A^2}-\frac{G_{33}}{C^2}-\frac{2G_{22}}{B^2},\\
R^{ab}R_{ab}&=&\frac{G_{00}^{2}}{A^4}+\frac{G_{33}^{2}}{C^4}+\frac{2G_{22}^{2}}{B^4}
-\frac{2G_{03}^{2}}{A^2C^2}.
\end{eqnarray*}
Thus the remaining part of the Weyl scalar becomes
\begin{eqnarray}\label{72}
\frac{1}{3}R^2-2R^{ab}R_{ab}&=&-\frac{5}{3}\frac{G_{00}^2}{A^4}-\frac{5}{3}\frac{G_{33}^2}{C^4}
-\frac{8}{3}\frac{G_{22}^{2}}{B^4}+\frac{4G_{03}^{2}}{A^2C^2}-
\frac{2}{3}\frac{G_{00}G_{33}}{A^2C^2}\nonumber\\
&+&\frac{4}{3}\frac{G_{22}G_{33}}{C^2B^2}-\frac{4}{3}\frac{G_{00}G_{22}}{A^2B^2}.
\end{eqnarray}
Using Eqs.(\ref{71}) and (\ref{72}), the Weyl scalar takes the
form
\begin{eqnarray}\label{73}
\mathcal{C}^{2}&=&\frac{48}{B^{6}}(m-\frac{s^2}{2B})^{2}-\frac{16}{B^3}(m-\frac{s^2}{2B})
[\frac{G_{00}}{A^2}-\frac{G_{33}}{C^2}+\frac{G_{22}}{B^{2}}]\nonumber\\
&-&\frac{4}{A^2C^2}G_{03}^{2}+3[(\frac{G_{00}}{A^2})^{2}+(\frac{G_{33}}{C^2})^{2}]
+\frac{4}{B^4}G_{22}^{2}-2\frac{G_{00}G_{33}}{A^2C^2}\nonumber\\
&+&4(\frac{G_{00}}{A^2}-\frac{G_{33}}{C^2})\frac{G_{22}}{B^2}
-\frac{5}{3}\frac{G_{00}^2}{A^4}-\frac{5}{3}\frac{G_{33}^2}{C^4}
-\frac{8}{3}\frac{G_{22}^{2}}{B^4}+\frac{4G_{03}^{2}}{A^2C^2}\nonumber\\
&-&\frac{2}{3}\frac{G_{00}G_{33}}{A^2C^2}+\frac{4}{3}\frac{G_{22}G_{33}}{C^2B^2}
-\frac{4}{3}\frac{G_{00}G_{22}}{A^2B^2}.
\end{eqnarray}

After some algebra, it leads to the following equation
\begin{equation}\label{74}
\frac{\mathcal{C}B^3}{(48)^\frac{1}{2}}=(m-\frac{s^2}{B})-\frac{B^3}{6}
[\frac{G_{00}}{A^2}+\frac{G_{22}}{B^2}-\frac{G_{33}}{C^2}-\frac{3s^2}{B^4}].
\end{equation}
Using the field equations, we have
\begin{equation}\label{75}
\frac{G_{00}}{A^2}+\frac{G_{22}}{B^2}-\frac{G_{33}}{C^2}-\frac{3\mu_{0}^{2}s^2}{B^4}
=8\pi(\mu+P_{\perp}-P_{z}+2\eta F).
\end{equation}
In view of the above equation and using $\tilde{Z}=B$,
Eq.(\ref{74}) becomes
\begin{equation}\label{76}
\frac{\mathcal{C}\tilde{Z}^3}{(48)^\frac{1}{2}}=[m-\frac{4\pi}{3}(\mu+P_{\perp}-P_{z}+2\eta
F)\tilde{Z}^3-\frac{s^2}{2\tilde{Z}}(\mu_{0}^{2}+1)].
\end{equation}
The derivatives of
$(\frac{\mathcal{C}\tilde{Z}^3}{(48)^\frac{1}{2}})$ with respect
to $\tilde{T}$ and $\tilde{Z}$ are given by
\begin{eqnarray}\label{77}
D_{\tilde{T}}(\frac{\mathcal{C}\tilde{Z}^3}{(48)^\frac{1}{2}})&=&-4\pi[\frac{1}{3}\tilde{Z}^{3}
D_{\tilde{T}}(\mu+P_{\perp}-P_{z}+2\eta
F)+(\mu+P_{\perp}+\epsilon+ \frac{2}{3}\eta F)\nonumber\\
&\times&\tilde{Z}^2U +(q+\epsilon)E\tilde{Z}^2]
+\frac{s^2U}{2\tilde{Z}^2}(\mu_{0}^{2}+1),\\\label{78}
D_{\tilde{Z}}(\frac{\mathcal{C}\tilde{Z}^3}{(48)^\frac{1}{2}})&=&4\pi
[(q+\epsilon)\frac{\tilde{Z}^2U}{E}-\frac{1}{3}\tilde{Z}^{2}
D_{\tilde{Z}}(\mu+P_{\perp}-P_{z}+2\eta
F)+(\epsilon-P_{\perp}\nonumber\\
&+&P_{z}-2\eta F)\tilde{Z}^2]
-\frac{sD_{\tilde{Z}}s}{\tilde{Z}}(\mu_{0}^{2}+1)+\frac{s^2}{2\tilde{Z^2}}(\mu_{0}^{2}+1).
\end{eqnarray}
These equations give relationship between the Weyl scalar and the
fluid properties like density, viscosity and pressure
(anisotropy). For perfect and non-charged fluid, Eq.(\ref{78})
reduces to the following form
\begin{equation}\label{79}
D_{\tilde{Z}}(\frac{\mathcal{C}\tilde{Z}^3}{(48)^\frac{1}{2}})=\frac{-4\pi}{3}\tilde{Z}^{3}
D_{\tilde{Z}}\mu.
\end{equation}
Using the regularity condition, it is concluded that
$D_{\tilde{Z}}\mu=0$ if and only if $\mathcal{C}=0$. This means
that if energy density is homogeneous, the metric is conformally
flat and vice versa.

We would like to mention here that a particularly simple relation
between the Weyl tensor and density inhomogeneity such as (2.76),
for perfect non-charged fluids, is at the origin of Penrose's
proposal to provide a gravitational arrow of time in terms of the
Weyl tensor \cite{38a}. The rationale behind this idea is that tidal
forces tend to make the gravitating fluid more inhomogeneous as the
evolution proceeds, thereby indicating the sense of time. However,
the fact that such a relationship is no longer valid in the presence
of local anisotropy of the pressure and/or dissipative processes
and/or electric charge. This has already been discussed \cite{25,
25a} explaining its failure in scenarios where the above-mentioned
factors are present \cite{25b}. Here we see how the electric charge
distribution affects the link between the Weyl tensor and density
inhomogeneity, suggesting that electric charge (whenever present)
should enter into any definition of a gravitational arrow of time.

\section{Dynamics of Viscous Adiabatic Case}

In this case, heat flux vanishes, also, we assume that radiation
density is zero and hence dissipation is only due to shearing
viscosity. The energy-momentum tensor is obtained by replacing
$q_{a}=\epsilon=0$ in Eq.(\ref{2}). Similarly, the Einstein-Maxwell
field equations are found by using $q=\epsilon=0$ in the
corresponding equations derived for non-adiabatic case. For junction
conditions, the line element for the exterior region is taken as
plane symmetric Reissner-Nordstr$\ddot{o}$m spacetime given by
\begin{equation}\label{80}
ds^{2}_{+}=-(\frac{-2M}{Z}+\frac{e^{2}}{Z^{2}})dT^{2}+(\frac{-2M}{Z}
+\frac{e^{2}}{Z^{2}})^{-1}dZ^{2}+Z^{2}(dX^2+dY^2),
\end{equation}
where $(\chi^{+0},\chi^{+1},\chi^{+2},\chi^{+3})=(T,X,Y,Z)$. The
equation of hypersurface in terms of exterior coordinates is
\begin{equation}\label{81}
k^{+}(T,Z)=Z-Z_{\Sigma}(T)=0.
\end{equation}
Using Eq.(\ref{81}), the exterior metric over the hypersurface
becomes
\begin{eqnarray}\label{82}
(ds^{2}_{+})_{\Sigma}&=&-[(\frac{-2M}{Z_{\Sigma}}+\frac{e^{2}}{Z_{\Sigma}^{2}})-
(\frac{-2M}{Z_{\Sigma}}+\frac{e^{2}}{Z_{\Sigma}^{2}})^{-1}(\frac{dZ_{\Sigma}}{dT})^{2}]dT^2\\\nonumber
&+&Z^{2}_{\Sigma}(dX^2+dY^2).
\end{eqnarray}

The first junction condition yields the following equations
\begin{eqnarray}\label{83}
\frac{dt}{d\tau}&=&\frac{1}{A}, \quad Z_{\Sigma}=B,
\\\label{85}
\frac{dT}{d\tau}&=&(\frac{-2M}{Z_{\Sigma}}+\frac{e^{2}}{Z_{\Sigma}^{2}})^{\frac{1}{2}}
[(\frac{-2M}{Z_{\Sigma}}+\frac{e^{2}}{Z_{\Sigma}^{2}})^{2}-
(\frac{dZ_{\Sigma}}{dT})^{2}]^{\frac{-1}{2}}.
\end{eqnarray}
Equation (\ref{85}) implies that
\begin{equation}\label{86}
d\tau^{2}=NdT^{2}-\frac{1}{N}dZ_{\Sigma}^{2},\quad
N=(\frac{-2M}{Z_{\Sigma}}+\frac{e^2}{Z_{\Sigma}^{2}}).
\end{equation}
The unit normal in terms of exterior coordinates is given by
\begin{equation}\label{88}
n^{+}_{a}=(-\hat{Z_{\Sigma}},0,0,\hat{T}),
\end{equation}
The non-null components of the extrinsic curvature for the
exterior spacetime are
\begin{equation}\label{89}
K_{00}^{+}=[\frac{dZ}{d\tau}\frac{d^{2}T}
{d\tau^{2}}-\frac{d^{2}Z}{d\tau^{2}}\frac{dT}{d\tau}-
\frac{N}{2}\frac{dN}{dZ}(\frac{dT}{d\tau})^{3}+
\frac{3}{2N}\frac{dN}{dZ}\frac{dT}{d\tau}(\frac{dZ}{d\tau})^{2}]_{\Sigma}.
\end{equation}
\begin{equation}\label{90}
K_{11}^{+}=[ZN \frac{dT}{d\tau}]_{\Sigma}=K_{22}^{+}.
\end{equation}

The second junction condition yields
\begin{equation}\label{91}
M\overset{\Sigma}{=}m(t,z)\quad \Leftrightarrow\quad
s\overset{\Sigma}{=}e.
\end{equation}
The rate of change of mass with respect to $\tilde{T}$ and
$\tilde{Z}$ are given by the following equations
\begin{equation}\label{92}
D_{\tilde{T}}m(t,z)=-4\pi(P_{z}-\frac{4}{3}\eta F)U\tilde{Z}^{2}
+\frac{s^{2}U}{2\tilde{Z}^{2}}(\mu_{0}^{2}-1),
\end{equation}
\begin{equation}\label{93}
D_{\tilde{Z}}m(t,z)=4\pi\mu
\tilde{Z}^{2}+\frac{s}{\tilde{Z}}D_{\tilde{Z}}s+
\frac{s^{2}}{2\tilde{Z}^{2}}(\mu_{0}^{2}-1).
\end{equation}
The description of these equations is the same as for the
non-adiabatic case. Similarly, the dynamical equations can be
obtained using $q=\epsilon=0$ in Eqs.(\ref{54})-(\ref{59}). For this
case, Eq.(\ref{59}) becomes
\begin{eqnarray}\label{94}
(\mu+P_{z}-\frac{4}{3}\eta
F)D_{\tilde{T}}U&=&-(\mu+P_{z}-\frac{4}{3}\eta
F)[\frac{m}{\tilde{Z}^{2}}+4\pi(P_{z}-\frac{4}{3}\eta
F)\tilde{Z}-\frac{s^{2}}{\tilde{Z}^{3}}]\nonumber\\
-E^{2}[D_{\tilde{Z}}(P_{z}-\frac{4}{3}\eta
F)&+&\frac{2}{\tilde{Z}}(P_{z}-P_{\perp}-2\eta
F)-\frac{\mu_{0}^{2}sD_{\tilde{Z}}s}{4\pi \tilde{Z}^{4}}].
\end{eqnarray}
The inequality given in Eq.(\ref{60}) remains the same. The
interpretation of the terms in Eq.(\ref{94}) is similar to that of
the non-adiabatic case just excluding the effects of heat flux and
radiation density. As there is no heat flux, so no transport
equation is needed. The relationship between the Weyl tensor and
energy homogeneity also remains the same.

\section{Conclusions}

Gravitational collapse is an outstanding phenomenon in gravitation
theory. The aim of this work is to analyze the dynamics of
gravitational collapse for plane symmetric configuration of real
fluid. The conclusions are given in the following
\begin{enumerate}
\item The junction conditions for both cases yield that
masses of the interior and exterior regions are equal if and only if
their corresponding charges are equal. For the non-adiabatic case,
junction conditions also give a relationship between heat flux,
effective pressure in $z$-direction and charge over the hypersurface
(\ref{43}). This equation implies that if $s=0$, then the effective
pressure in $z$-direction and heat flux are equal over the
hypersurface.
\item For the non-dissipative case, Eq.(\ref{48}) reduces to
$D_{\tilde{Z}}(\frac{U}{\tilde{Z}})=0$ which shows that collapse is
homologous, i.e., all the matter falls inward in a similar pattern.
\item Condition for hydrostatic equilibrium is
given by Eq.(\ref{61}). If this condition holds, then the collapsing
process will stop and matter attains an equilibrium state.
\item It is observed that charge will increase
the rate of collapse if $\frac{s}{\tilde{Z}}>D_{\tilde{Z}}s$. Thus
the chance of becoming a black plane increases in this case.
\item In the non-adiabatic case, the radiation density increases inertial and
active gravitational masses. Also, the outflow of radiation causes
an increase in the rate of collapse and hence the collapsing process
is expected to be faster than the adiabatic case.

\item The substitution of transport equation (\ref{63}) in
dynamical equation (\ref{59}) yields an additional factor $\alpha$.
This $\alpha$ term affects the inertial mass density and
gravitational force term. As $\alpha$ increases, the affected terms
are decreased by the same amount and vice versa. The corresponding
terms for $\alpha$ come from the term $a_{b}T$ in Eq.(\ref{62})
which is the Tolman's inertial term. Hence the inertia of heat by
increasing $\alpha$, causes a decrease in inertial mass and
gravitational force \cite{38}. Thus we can conclude that
\begin{itemize}
\item If $\alpha\rightarrow0$, then inertial density and gravitational
force are not affected by coupling.
\item If $0<\alpha<1$, then inertia of heat causes a decrease in inertial
and gravitational mass densities.
\item If $\alpha\rightarrow1$, then mass densities approach to zero.
\item If $\alpha>1$, then the gravitational force term becomes negative
implying that reversal of collapse occurs due to the inertia of
heat.
\end{itemize}
\item Under certain conditions homogeneity in energy density and conformal
flatness of spacetime are necessary and sufficient condition for
each other.
\item A relation (\ref{78}) has been obtained exhibiting the way in which
electric charge affects the link between the Weyl tensor and density
inhomogeneity.
\end{enumerate}

\end{document}